\documentclass[final,5p,times,twocolumn,authoryear]{elsarticle}

\usepackage{amssymb}
\usepackage{amsmath}
\usepackage[flushleft]{threeparttable}
\usepackage{url}
\usepackage{multirow}
\usepackage{xcolor}

\journal{Journal of High Energy Astrophysics}

\begin{document}

\begin{frontmatter}



% \title{Spectral Energy Distribution Analysis of CAL\,87 with \textit{XMM--Newton} and Multiwavelength Data} %% Article title
\title{Spectral Energy Distribution Analysis of the Supersoft X-ray Source CAL\,87: Multiwavelength Constraints}


% \author{} %% Author name

\author[a,b,c]{Paulo E. Stecchini}
\author[c]{Francisco Jablonski}
\author[b]{Marcos P. Diaz}
\author[a]{Alexandre S. Oliveira}
\author[c]{Flavio D'Amico}
\author[a]{Natália Palivanas}

\affiliation[a]{
organization={IP\&D, Universidade do Vale do Paraíba}, adressline={Av. Shishima Hifumi, 2911,}, city={São José dos Campos}, postcode={12244-000},state={SP},country={Brazil} 
}
\affiliation[b]{
organization={Instituto de Astronomia, Geofísica e Ciências Atmosféricas, Universidade de São Paulo}, adressline={Rua do Matão, 1226,}, city={São Paulo}, postcode={05508–900},state={SP},country={Brazil} 
}
\affiliation[c]{
organization={Divisão de Astrofísica, Coordenação de Engenharia, Tecnologia e Ciências Espaciais, Instituto Nacional de Pesquisas Espaciais}, adressline={Av. dos Astronautas, 1758,}, city={São José dos Campos}, postcode={12227-010},state={SP},country={Brazil} 
}

\begin{abstract}
We analyse the spectral energy distribution (SED) of the eclipsing supersoft X-ray source CAL\,87 covering wavelengths from X-rays to the near-infrared. Our study incorporates 26 data points across ultraviolet to near-infrared, sourced from published literature, unpublished data, and new observations. In addition, archival \textit{XMM$-$Newton} spectra were used to represent the X-ray emission. Care was taken to use out-of-eclipse flux measurements when the irradiated side of the companion faces the observer. The SED model includes contributions from a central source, a reprocessed accretion disk, and an irradiated companion star atmosphere, resulting in a good match to the observed fluxes. The revised and new parameters for the disk and the central source align with previous studies and match expectations for such systems. The temperature of the irradiated side of the companion star was estimated based on its $B-V$ colour during the secondary eclipse. This work highlights the importance of broad wavelength coverage for understanding the properties of supersoft X-ray sources.
\end{abstract}


% \begin{highlights}
% \end{highlights}

%% Keywords

\end{frontmatter}



\section{Introduction}\label{sec:intro}
The Supersoft X-ray Sources (SSS) are binary stellar systems first detected in the Large Magellanic Cloud (LMC) by the \textit{Einstein Observatory} \citep{1981ApJ...248..925L} and later defined as a distinct class through observations with the \textit{Röntgen Satellite} (\textit{ROSAT}) \citep{1991Natur.349..579T}. These systems are characterised by their exceptionally soft X-ray spectra (typically below 1\,keV) and near-Eddington luminosities (e.g. \citealp{1997ARA&A..35...69K}), which originate from steady nuclear burning on the surface of the white dwarf (WD) component \citep{1992A&A...262...97V}. This nuclear burning is sustained by very high accretion rate regimes ($\dot{M}\,\sim\,10^{-7}$\,M$_{\odot}$\,yr$^{-1}$), which may occur if the donor star is more massive than the WD (dynamical instability mass transfer model, DIMT; \citealt{1992A&A...262...97V}) or due to mass transfer driven by the wind of a strongly irradiated low-mass donor star (wind-driven mass transfer model, WDMT; \citealt{1998A&A...338..957V}). By now, several SSS are known in the LMC, Small Magellanic Cloud (SMC), M31, and other galaxies, with a few in the Milky Way (e.g. \citealp{1997ARA&A..35...69K}; \citealp{1998PASP..110..276S}; \citealp{2000NewA....5..137G}; \citealp{2021A&A...646A..85G}).

CAL\,87 is one of the prototypes of the SSS class. Located in the LMC, it is an eclipsing system with an orbital period of 10.6\,h \citep{1989MNRAS.241P..37C}. The optical light curve shows primary and secondary asymmetric eclipses. Models of the light curve \citep{1997A&A...318...73S, 1997A&A...321..245M} attribute this asymmetry and the depth of the eclipses to an irradiated secondary star in a high-inclination (77$^{\circ}-78^{\circ}$) binary. They also identify the dominant source of optical light as a large, optically thick spray at the rim of the irradiated accretion disk. This elevated structure, formed as the high-rate accretion flow collides with the disk, extends azimuthally and acts as a screen for the hot WD radiation, consistently blocking a direct view of the WD. This geometric configuration has important consequences for the observed X-ray emission, which is thought to originate not from the WD photosphere itself, but rather from reprocessed or scattered radiation emerging from an extended accretion disk corona (ADC) or disk wind, implying that the intrinsic luminosity of the WD must be significantly higher than that directly inferred from spectral fits (e.g. \citealp{1997A&A...322..591H}; \citealp{2004RMxAC..20...18G}; \citealp{2001ApJ...550.1007E}; \citealp{2024ApJ...960...46T}).

An early analysis of the spectral energy distribution (SED) in SSS was performed by \cite{1996LNP...472...65P}, who calculated SED models and compared the resulting fluxes to ultraviolet (UV) and optical observations of three SSS, including CAL\,87. Their main conclusion was that, since steady nuclear burning at the WD surface greatly exceeds the disk's accretion luminosity, the primary role of the disk is to reprocess the radiation from the central source. These SED models include the spectrum of the WD, a flared disk, and an illuminated secondary star, fitting the UV and optical fluxes reasonably well. They noted a strong dependence of the predicted optical flux on the size of the disk and, particularly for CAL\,87, the influence of the system's inclination in reducing the optical flux.

In this paper, we present an analysis of the spectral energy distribution of CAL\,87, covering wavelengths from X-rays to the near-infrared (NIR). Our study utilises X-ray spectra from the archival \textit{XMM$-$Newton} observation---used to represent the soft X-ray band in the SED---along with a comprehensive set of additional measurements, including published and unpublished data, as well as new observations, totalling 26 data points at longer wavelengths. We apply an irradiated disk model to represent the system's disk and an atmosphere model for the companion star, the latter based on the temperature derived from the colours observed during secondary eclipses. These models are integrated into our fitting process, with a multiplicative scaling factor used to adjust the atmosphere model as needed. In our model, the luminosity of the central irradiating source that drives the disk emission is treated as an independent parameter, rather than being derived from the observed X-ray flux alone, reflecting the obscured nature of the WD and guided by values suggested in the literature.

\section{Observations and flux measurements} \label{sec:data}

\subsection{X-rays}\label{sec:dataxray}

CAL\,87 was observed by the \textit{XMM$-$Newton} mission \citep{2001A&A...365L...1J} on a single occasion (ObsID\,0153250101). The science products derived from the X-ray instruments have been analysed from multiple angles: continuum fits with \mbox{EPIC-pn} and \mbox{EPIC-MOS\,1,2} \citep{2010AN....331..152E}, high-resolution line studies with \mbox{RGS\,1,2} (e.g. \citealp{2015ApJ...815...17A}; \citealp{2024ApJ...960...46T};  \citealp{2024A&A...690A...9P}), and timing analyses (e.g. \citealp{2014ApJ...792...20R}; \citealp{2024MNRAS.527.8991S}). 

Although our SED model (Section\,\ref{sec:sed}) treats the central irradiating source independently, we include a soft X-ray representation to maintain continuity across the spectral energy distribution. To that end, we fit the high-resolution RGS spectra jointly with EPIC-pn: the gratings resolve the prominent emission-line structure in the soft band, while pn adds high throughput and sensitivity at very soft energies (down to $\approx$\,0.2\,keV), improving the signal-to-noise ratio and extending the usable bandpass.

Data reduction was conducted with the \textit{XMM$-$Newton} Science Analysis System (SAS, v.\,20.0.0), following standard procedures. The observation spanned two orbital cycles; we selected off-eclipse good-time intervals to build the spectra. Consistent with the source softness and instrument calibration, we restricted the fits to \mbox{0.2--1.0}\,keV (pn) and 0.35--0.95 \,keV (RGS). RGS1 and RGS2 spectra were combined with the SAS task \texttt{rgscombine};  data were grouped to a minimum of 10 counts per bin for $\chi^2$ fitting in \textsc{XSPEC} \citep[][v.\,12.14.0]{1996ASPC..101...17A}. 

Previous analyses have identified the main spectral signatures for CAL\,87 in this energy range. The oxygen photoelectric edges near O\,\textsc{vii} (0.739\,keV) and O\,\textsc{viii} (0.871\,keV) were already noted in moderate-resolution \textit{ASCA}/SIS spectra \citep{1998ApJ...503L.143A,2001ApJ...550.1007E}, and high-resolution grating studies with \textit{XMM$-$Newton}/RGS and \textit{Chandra}/LETG further established that the $\sim$\,0.5–0.8\,keV band is line-dominated (e.g. \citealp{2010AN....331..152E}; \citealp{2024ApJ...960...46T}; \citealp{2024A&A...690A...9P}). 

\begin{figure}
% \centering
	\includegraphics[width=\columnwidth]{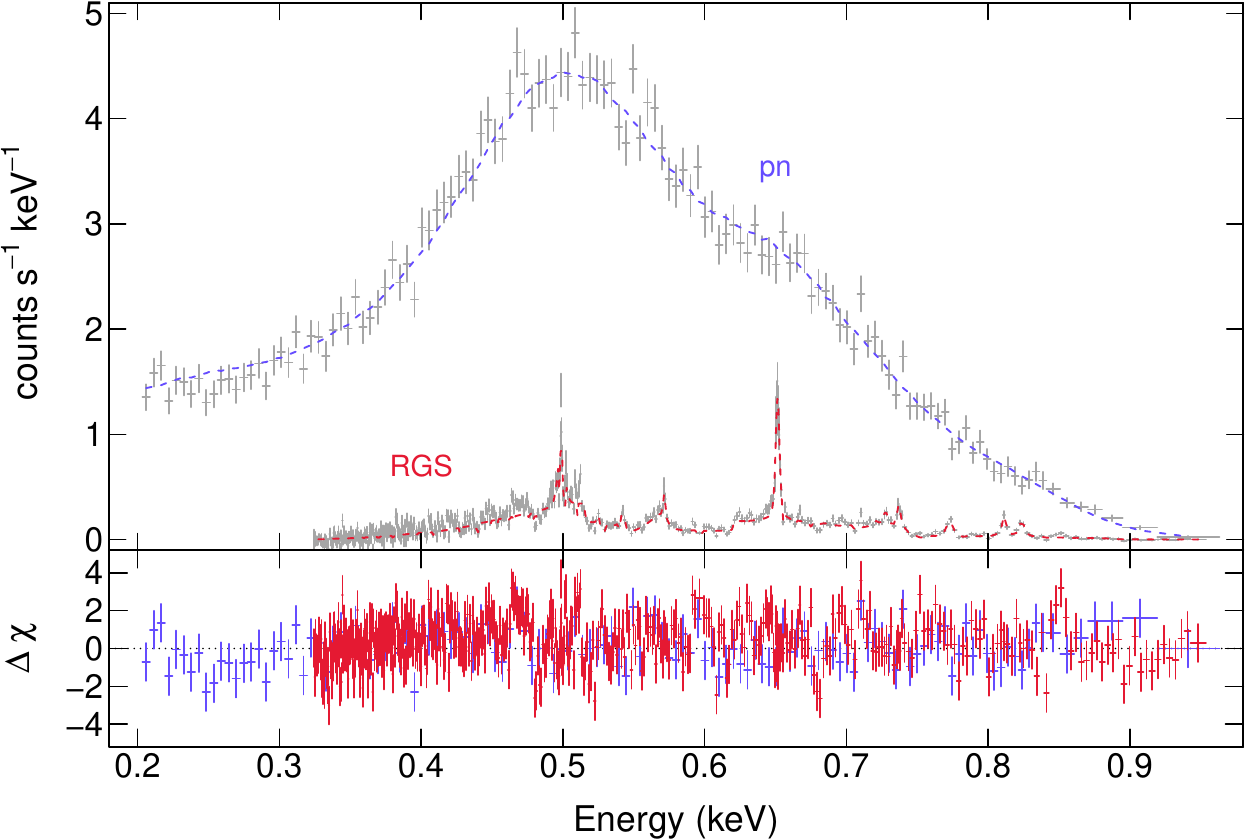}
    \caption{Joint RGS+pn fit with an absorbed blackbody plus two edges and 12 Gaussian lines. Residuals are shown in the lower panel.}
    \label{fig:xray}
\end{figure}

Motivated by these features, we model the joint RGS+pn data as an absorbed blackbody with two absorption edges and 12 Gaussian lines. The edges approximate the O\,\textsc{vii}/O\,\textsc{viii} thresholds, and the emission lines are represented by narrow Gaussians with centroids fixed to the measured values reported in table~1 of \cite{2024A&A...690A...9P}. A multiplicative constant is included to account for cross-calibration. In \textsc{XSPEC} notation this corresponds to \texttt{constant*tbabs*edge*edge*(bbody\,+\,$\sum$\,gaussian)}. Best-fitting parameters are listed in Table\,\ref{tab:01}, and the spectra, model and residuals are shown in Figure\,\ref{fig:xray}. 

\begin{table}
\caption{RGS+pn fit with \texttt{edge*edge*(bbody+$\sum$\,gaussian)}.\label{tab:01}}
\centering
\begin{threeparttable}
{\renewcommand{\arraystretch}{1.25}
\begin{tabular}{l c}
\hline\hline
Parameter & Best-fitting value \\  
\hline
$N_{\text{H}}$ [10$^{22}$\,cm$^{-2}$]   & 0.37$\,\pm$\,0.03           \\
$kT$ [eV]                               & 72\,$\pm$\,3             \\
$E_{\rm edge}$\,(O\,\textsc{viii}) [keV]                 & 0.881\,$\pm$\,0.011              \\
$\tau_{\rm edge}$\,(O\,\textsc{viii})                    & 3.43$^{+0.00}_{-1.31}$            \\
$E_{\rm edge}$\,(O\,\textsc{vii}) [keV]                 & 0.742\,$\pm$\,0.003              \\
$\tau_{\rm edge}$\,(O\,\textsc{vii})                    & 1.32$^{+0.30}_{-0.20}$              \\
$F_{\rm{abs}}$ [10$^{-12}$\,erg\,s$^{-1}$\,cm$^{-2}$] & 2.2\,$\pm$\,0.2              \\
$\chi^2_{\rm{red}}\,(\rm{d.o.f.})$           & 1.10 (1337)     \\ \hline
\hline
\end{tabular}
}
\begin{tablenotes}
\item \small \textbf{Notes.} The model includes a constant for cross-normalisation and the absorption model \texttt{tbabs}. Line centroids are fixed to the values in table~1 of \cite{2024A&A...690A...9P}. $F_{\rm abs}$ refers to the absorbed flux in the 0.2--1.0\,keV band and is computed with command \texttt{cflux} in \textsc{XSPEC}. Uncertainties are 90\%.
\end{tablenotes}
\end{threeparttable}
\end{table}

Overall fit parameters are consistent with previous work: the blackbody temperature (69--75\,eV) and the column density (0.34--0.40\,$\times$\,10$^{22}$\,cm$^{-2}$) lie within published ranges, the latter also compatible with the line-of-sight value toward the LMC of 0.34\,$\times$\,10$^{22}$\,cm$^{-2}$ \citep{2016A&A...594A.116H}, allowing for local absorption. The absorbed 0.2--1.0\,keV flux is $F_{\text{abs}}$\,$\sim$\,2.2\,$\times$\,10$^{-12}$\,erg\,cm$^{-2}$ s$^{-1}$; at $\sim$50\,kpc this corresponds to an unabsorbed luminosity $L_{0.2-1.0}$\,$\sim$\,1--2\,$\times$\,10$^{37}$\,erg\,s$^{-1}$. For context, \citet{2024ApJ...960...46T} obtain for RGS+MOS a comparable luminosity with similar phenomenological model, whereas \citet{2024A&A...690A...9P} report $\sim$\,5$\times$\,10$^{36}$\,erg\,s$^{-1}$ for LETG or RGS from a more physically motivated model (hot atmosphere plus optically thin plasma). Variations in unabsorbed values may reflect differences in the $N_{\text{H}}$ inferred by the spectral model, and we do not attempt to reconcile them here.

In interpreting these numbers, it is worth recalling that the observed soft X-rays in CAL\,87 are widely attributed to scattered or reprocessed emission---likely from an extended corona or disk wind---rather than direct radiation from the white dwarf (e.g. \citealp{2010AN....331..152E}; \citealp{2024A&A...690A...9P}). Accordingly, luminosities from the X-ray fit describe the observed scattered component and do not directly constrain the intrinsic luminosity of the central source. For the SED, we use the unabsorbed RGS+pn model to define the soft-X spectral shape, and fit the irradiating luminosity independently (Section~\ref{wd}).

\subsection{Longer Wavelengths}

Flux measurements in longer wavelengths were either sourced directly from the literature or determined by us. For the latter, the fluxes we determined came from publicly available data that had not yet been analysed for the same purpose, as well as from new observations. 

Previously reported measurements of CAL\,87 include data from instruments of the Hubble Space Telescope (HST), the Cerro Tololo Inter-American Observatory (CTIO), the Optical Gravitational Lensing Experiment (OGLE), the Massive Compact Halo Objects project (MACHO), and the Visual and Infrared Telescope for Astronomy (VISTA) Magellanic Clouds survey (VMC). References for these measurements are provided in the notes to \hbox{Table\,\ref{tab:02}}; unpublished and new data will be explained shortly. In some cases, flux values originate from photometry taken during a specific orbital phase of the source, thereby limiting the measurement to that phase, or from light curves that allowed us to select intervals of interest (primarily outside of the eclipses). In \hbox{Table\,\ref{tab:02}}, we employed different notations to represent these situations: a phase span, marked by an en dash, signifies that photometry was conducted continuously over that specified range; multiple phases separated by commas denote that the fluxes were measured at each of these phases, and the reported value is their average. Finally, ``OoE" stands for ``out of eclipse" and refers to the average flux derived from a well-sampled light curve, measured outside the eclipse phases. 

Publicly available but unpublished measurements were obtained from the Galaxy Evolution Explorer (GALEX; \citealp{2005ApJ...619L...1M}), the Gaia Satellite Mission (Gaia; e.g. \citealp{2022gdr3.reptE....V}) and the Optical Monitor (OM; \citealp{2001A&A...365L..36M}) on board the \textit{XMM$-$Newton} mission. The GALEX and Gaia measurements were sourced directly from the Mikulski Archive for Space Telescopes\footnote{\url{https://mast.stsci.edu/} \citep{https://doi.org/10.17909/t9h59d}.} (MAST) and the Gaia ESA Archive\footnote{\url{https://gea.esac.esa.int/archive/} \citep{gaiadr3}.}, respectively. The OM measurements were obtained from aperture photometry on the images available from the same X-ray observation. For the U and V filters, calibration was performed using a set of isolated objects with available photometry data from the APASS DR10 catalog\footnote{\url{https://www.aavso.org/download-apass-data}.}. For the M2 and W1 filters, we used the SAS tasks for flux conversion factors, as described in their documentation page\footnote{\url{https://xmm-tools.cosmos.esa.int/external/xmm_user_support/documentation/sas_usg/USG/ommag.html}.}.

Additionally, new {\em griz} measurements were obtained on November 6, 2023, with the SPARC4\footnote{SPARC4 is a photometry/polarimetry instrument that takes simultaneous images in the Sloan {\em griz} bands.}  \citep{2012AIPC.1429..252R} camera installed at the 1.6-m telescope of Pico dos Dias Observatory in southeast Brazil\footnote{\url{https://www.lna.br}.}. In this case, photometric calibration was performed using 13 isolated objects from the SkyMapper Southern Sky Survey\footnote{\url{https://skymapper.anu.edu.au/}.}.

We also retrieved light curves from the European Southern Observatory (ESO) in the B band \citep{1989MNRAS.241P..37C} and MACHO, in the V band\footnote{\url{http://cdsarc.u-strasbg.fr/viz-bin/VizieR-5?-ref=VIZ642dda8b315c&-out.add=.&-source=II/247/machovar&recno=4458} \citep{vizier}.}, to examine the colours during the secondary eclipse and discuss the companion star in the system. 

The uncertainties listed in Table\,\ref{tab:02} refer solely to factors inherent in the measurements, such as instrumental or photometry errors. Those not displayed stem from missing information in the original publication. 

\begin{table*}
\caption{Measurements of CAL\,87 in longer wavelengths.\label{tab:02}}
\centering
\begin{threeparttable}
{\renewcommand{\arraystretch}{1.25}
\begin{tabular}{c c c c c}
\hline\hline
Wavelength & Flux & Uncertainty & \multirow{2}{*}{Phase} & \multirow{2}{*}{Id}\\ 
(Angstroms) & ($10^{-16}$\,erg\,cm$^{-2}$\,s$^{-1}$\,\AA$^{-1}$) & \%\\
\hline
1775        & 6.350                                              & -           & 0.20, 0.67             & HST/UV$^{\rm{a}}$                          \\
2275        & 4.190                                              & 13.22       & 0.28                   & GALEX/NUV                       \\
2310        & 4.120                                              & 03.22       & 0.13                   & OM/M2                           \\
2910        & 3.270                                              & 03.24       & 0.74                   & OM/W1                           \\
3440        & 3.200                                              & 02.49       & 0.35                   & OM/U                            \\
3600        & 2.690                                              & -           & 0.30--0.36             & HST/\textit{U}$^{\rm{b}}$                           \\
4300        & 1.780                                              & -           & 0.30--0.36             & HST/\textit{B}$^{\rm{b}}$                           \\
4300        & 1.680                                              & -           & OoE                    & CTIO/\textit{B}$^{\rm{c}}$                          \\
4702        & 1.542                                              & 00.47       & 0.56--0.64             & SPARC4/{\em g}                  \\
5124        & 1.361                                              & 09.21       & OoE                    & Gaia/Gbp                                 \\
5430        & 1.260                                              & 06.54       & 0.68                   & OM/V                            \\
5500        & 1.060                                              & -           & 0.30--0.36             & HST/\textit{V}$^{\rm{b}}$                           \\
5500        & 1.070                                              & 00.39       & OoE                    & OGLE/\textit{V}$^{\rm{e}}$                          \\
5500        & 1.080                                              & -           & OoE                    & CTIO/\textit{V}$^{\rm{d}}$                          \\
6176        & 0.981                                              & 00.46       & 0.56--0.64             & SPARC4/{\em r}                  \\
6252        & 0.688                                              & 04.61       & OoE                    & Gaia/G                                   \\
7000        & 0.580                                              & -           & 0.30--0.36             & HST/\textit{R}$^{\rm{b}}$                           \\
7000        & 0.544                                              & -           & OoE                    & CTIO/\textit{R}$^{\rm{d}}$                          \\
7490        & 0.614                                              & 00.66       & 0.56--0.64             & SPARC4/{\em i}                  \\
7830        & 0.492                                              & 09.21       & OoE                    & Gaia/Grp                                 \\
8947        & 0.429                                              & 01.51       & 0.56--0.64             & SPARC4/{\em z}                  \\
9000        & 0.323                                              & -           & OoE                    & CTIO/\textit{I}$^{\rm{d}}$                          \\
9000        & 0.322                                              & 00.36       & OoE                    & OGLE/\textit{I}$^{\rm{e}}$                          \\
10200       & 0.276                                              & 01.31       & 0.25, 0.33, 0.43       & VMC/\textit{Y}$^{\rm{e}}$                           \\
12350       & 0.152                                              & 02.16       & 0.23, 0.58             & VMC/\textit{J}$^{\rm{e}}$                           \\
21590       & 0.032                                              & 02.37       & OoE                    & VMC/\textit{K}$_{\textrm{\scriptsize S}}$$^{\rm{e}}$ \\
\hline
\end{tabular}
}
\begin{tablenotes}
\item \small \textbf{Notes.} The first column lists the central wavelengths, while the second column shows the observed fluxes, i.e. without any correction for extinction. Uncertainties account only for instrumental and/or photometry errors. ``OoE" stands for ``out of eclipse". Further details can be found in the text.
\item
\textbf{References.}
$^{\rm{a}}$\hbox{\cite{1995AJ....110.2394H}}; $^{\rm{b}}$\hbox{\cite{1996ApJ...471..979D}}; $^{\rm{c}}$\hbox{\cite{1990ApJ...350..288C}}; $^{\rm{d}}$\hbox{\cite{1991ApJ...373..228C}}; $^{\rm{e}}$\hbox{\cite{2024MNRAS.527.8991S}}.
\end{tablenotes}
\end{threeparttable}
\end{table*}

\section{The Spectral Energy Distribution of CAL\,87} \label{sec:sed}

In this section, we present the analysis of the spectral energy distribution of CAL\,87. We begin by explaining how the measurements were corrected for optical extinction, followed by a discussion of each component involved in constructing the SED. Finally, we describe the fitting process, with the Markov Chain Monte Carlo (MCMC) analysis concluding the section. The best-fitting parameters, obtained from the posterior distribution in the MCMC analysis, are listed in the last column of Table\,\ref{tab:03}; the resulting SED, including the model components based on these posteriors, is shown in Figure\,\ref{fig:sed}.

\subsection{The Optical Extinction}\label{av}

Two methods were employed to estimate the optical extinction $A_{\text{V}}$ along the line of sight of CAL\,87. The first involved directly reading from figure\,9 (middle panel) of \cite{2021MNRAS.508..245M}, which presents $V$-band extinction maps derived from NIR photometry of the VMC. Using the coordinates of CAL\,87, we found a value of $A_{\text{V}}$\,=\,0.34\,$\pm$\,0.1\,mag, with the estimated uncertainty based on the colour coding of neighboring tiles. The second method is based on the map produced by the reddening tool from results of the OGLE-IV survey in the LMC. The reddening values are provided in terms of $E(V-I)$, which, for the region of CAL\,87, has a median of 0.153$^{+0.065}_{-0.056}$, with 1$\sigma$ confidence level. The transformation of $E(V-I)$ to $A_{\text{V}}$ was done using the extinction law of \cite{2014A&A...564A..63M}. An interesting aspect of that work is that the extinction laws were derived from monochromatic measurements, thus avoiding the complications of integrating inside photometric passbands. From the pivotal wavelengths of OGLE-IV $V$ and $I$ bands, we found that a value of $A_{\text{V}}$\,=\,0.335 was required to produced the observed $E(V-I)$\,=\,0.153\,mag. Using the complete response of the OGLE-IV $V$ and $I$ filters yields $A_{\text{V}}$\,=\,0.338$^{+0.144}_{-0.123}$, which are the value and uncertainty we adopted. The measurements displayed in Figure\,\ref{fig:sed} are already corrected for this value.  

\subsection{The White Dwarf and Irradiating Source}\label{wd}

Hydrogen burning in hydrostatic equilibrium beneath the white dwarf photosphere, at near-Eddington luminosities, is believed to be the main source of soft X-rays in high-state SSS. As discussed in Section\,\ref{sec:dataxray}, however, the soft X-ray emission in CAL\,87 is likely not observed directly, but instead arises from reprocessed or scattered radiation. For this reason, the observed X-ray spectrum cannot be used to constrain parameters such as the WD’s radius or mass in a straightforward manner. Our spectral fit provides an estimate of the observed flux, which guides the choice of luminosity for the central irradiating source in the irradiated disk model.

Convective energy transfer and optically thick winds may lead to a pseudo-photosphere radius that is much larger than the unperturbed radius of a WD with the same inner structure. Observations of novae during their supersoft phase support this, often implying photospheric radii well above theoretical mass--radius expectations, even for low-mass WDs \citep{2020MNRAS.499.4814P}. For this reason, the mass--radius relation from \citet{1972ApJ...175..417N} provides only lower limits to the effective radius contributing to the observed SED. Although the effective radius of the photosphere is not required for the SED fitting, the actual WD radius is, and we allow it to vary independently of the mass within the fitting process. A flat prior on the WD mass was adopted, ranging from 0.6 to 1.4 M$_{\odot}$.

To distinguish the properties of the irradiating source used in the SED model from those derived in the X-ray spectral fit, we adopt the subscript “irr” for its mass ($M_{\rm{irr}}$), radius ($R_{\rm{irr}}$), and luminosity ($L_{\rm{irr}}$). While this source is presumed to be associated with the accreting WD, its parameters are determined independently through the UV/optical/IR SED fit. The prior lower limit on $L_{\rm{irr}}$ was guided by the unabsorbed luminosity inferred from our X-ray spectral analysis, while acknowledging that the disk may intercept a significant fraction of the WD’s intrinsic output even if the observer does not. The allowed range was set wide enough to encompass lower luminosities, such as those inferred under smaller absorbing column densities in other studies, as well as higher values, allowing for the possibility of additional contributors to the irradiation, such as the boundary layer or innermost disk. 

\subsection{The Accretion Disk}\label{disk}

The accretion disk spectrum is calculated from an accretion disk with central source irradiation, as described by \citet{1998PASJ...50...89M}. A flared, optically thick disk is supported against the vertical component of gravity by gas and radiation pressure. Incident radiation from the central source is thermalized at the disk surface, with a homogeneous albedo $A$. In this model, the two-component radial temperature profile is suitable for irradiated and vertically extended disks, where the outer disk has an extended aspect seen from the central source. The temperature follows a power-law distribution, transitioning from $T$\,$\propto$\,$r^{-3/4}$ in the inner disk to $T$\,$\propto$\,$r^{-3/7}$ in the outer region due to irradiation and flaring effects. The viscous dissipation of the disk is not considered (see Section\,\ref{sec:discussion}). The total disk spectrum is computed as the integrated local blackbody emission from the innermost radius $R_{\text{in}}$ to the outer disk radius $R_{\text{out}}$, with azimuthal symmetry. Both the inner and outer radii are allowed to vary freely: the inner radius is linked to the white dwarf radius, which depends on the fit, while the outer radius has limits around a fraction (80\%) of the primary Roche lobe. The disk inclination, also a free parameter, is expected to be near $i$\,=\,70$^{\circ}$ for CAL\,87 \citep[e.g.][]{1997A&A...318...73S}. In our SED fitting, the disk model includes seven free parameters in total: the inner and outer radii ($R_{\text{in}}$, $R_{\text{out}}$), inclination angle ($i$), and albedo ($A$), along with the irradiating source's luminosity ($L_{\rm{irr}}$), radius ($R_{\rm{irr}}$) and mass ($M_{\rm{irr}}$)---the latter three discussed in the previous section. These parameters are listed in Table\,\ref{tab:03}.  

\subsection{The Secondary}

\begin{figure}
% \centering
	\includegraphics[width=\columnwidth]{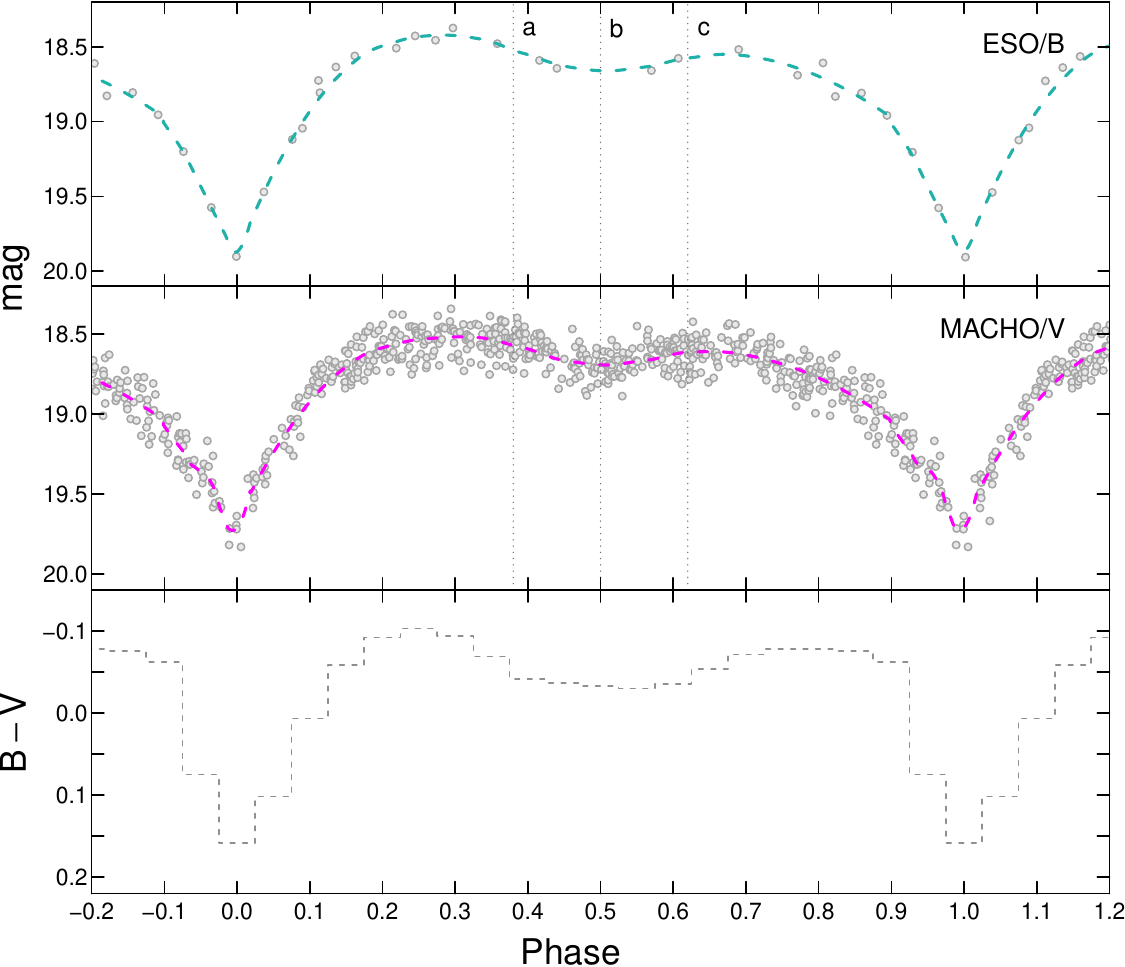}
    \caption{Folded light curves of CAL\,87 from ESO (B band, upper panel) and MACHO (V band, middle panel) observations. A spline fit is superimposed on the data points. Orbital phases (\textit{a}, \textit{b}, and \textit{c}) used in the computation of the secondary eclipse $B-V$ index (equation\,\ref{eq:01}) are indicated. Bottom panel shows the $B-V$ index, with binning similar to that of figure 2 in \cite{1989MNRAS.241P..37C}.}
    \label{fig:b-v}
\end{figure}

Figure\,\ref{fig:b-v} shows the folded light curves of CAL\,87 in the $B$ and $V$ bands, according to the ephemeris of \cite{2024MNRAS.527.8991S}. The magnitudes for both curves are corrected for reddening (Section\,\ref{av}). The $B-V$ index is also shown, in the bottom panel.
We posit that a fraction of the secondary star's photosphere, at least partially illuminated by the white dwarf and inner accretion disk, faces the observer just before point \textit{a} and after point \textit{c}, while being significantly eclipsed (point \textit{b}) by the accretion disk between points \textit{a} and \textit{c}. A change in colour index during these orbital phases could indicate the colour of the eclipsed secondary, as the observed SED becomes bluer or redder during its occultation. At point \textit{b}, during the secondary eclipse, we observe less of the companion star surface. To determine the colour of this eclipsed component, magnitudes must be converted into flux, since the occultation-induced loss is additive to the flux level outside of the eclipse:

\begin{align}\label{eq:01}
(B-V)_{\rm{ecl}} = -2.5\,\rm{log_{10}} \left[10^{-0.4\,B_{\rm{ac}}} - 10^{-0.4\,B_{\rm{b}}}\right]\nonumber \\ + 2.5\,\rm{log_{10}} \left[10^{-0.4\,V_{\rm{ac}}} - 10^{-0.4\,V_{\rm{b}}}\right]
\end{align}

\noindent where B$_{\rm{ac}}$ and V$_{\rm{ac}}$ represent the mean magnitudes in each band for points $a$ and $c$ (outside the eclipse), and B$_{\rm{b}}$ and V$_{\rm{b}}$ are the magnitudes in each band for point $b$ (at the minimum of the eclipse).
We assume that the differential flux between the two sampled phases within the secondary eclipse is due to the geometric occultation effect only, neglecting other sources of variability in the system during the eclipse. The resulting ($B-V$) colour of the eclipsed component should be bluer than that seen at the opposite companion hemisphere, which is not affected by irradiation from the WD. However, the secondary eclipse colour may be biased by the different aspect of the eclipsing disk itself, which lacks azimuthal symmetry, as seen between points \textit{a} or \textit{b} and point \textit{c}. Neglecting the SED variation caused by the different view of the disk, we estimate, using equation\,\ref{eq:01} (and B$_{\rm{ac}}$\,=\,18.58, B$_{\rm{b}}$\,=\,18.67, V$_{\rm{ac}}$\,=\,18.58, V$_{\rm{b}}$\,=\,18.71), a $B-V$ index of approximately 0.36, which corresponds,  according to  \cite{2013ApJS..208....9P}, to a temperature of the heated secondary of 7000\,$\pm$\,1000\,K. An irradiated inner face is thus assumed for the companion star, as expected from the system geometry and the likely presence of an extended scattering medium that redirects central source radiation onto the donor’s surface (e.g. \citealp{1997A&A...318...73S}).

The corresponding SED component was calculated as a solar-abundance, log($g$)\,=\,5.0, non-irradiated stellar atmosphere using the \texttt{TLUSTY} \citep{hubeny1988, tlusty2011} and \texttt{SYNSPEC} \citep{synspec2011} codes. In our fitting process, the SED of the secondary is parameterized solely by an arbitrary overall scaling factor---denoted ``Atmosphere factor" in Table\,\ref{tab:03}---which is intended to compensate for the simplified description of the companion-irradiation process.

\subsection{Markov Chain Monte Carlo (MCMC) for exploration of parameters}
We used our own version of the MCMC procedure to explore the space of parameters of the disk model by \citet{1998PASJ...50...89M} and the irradiated companion level (``Atmosphere factor"). The MCMC algorithm is particularly suited for complex models with non-analytic outputs. The goal is to access the posterior distribution of the model parameters, $P(\theta|D)$, by sampling different sets of parameters that produce varying likelihoods of the data, $P(D|\theta)$, while accounting for any prior knowledge of the parameters, $P(\theta)$.

The MCMC algorithm is relatively simple to code, and we refer the reader to the original work by \cite{1953JChPh..21.1087M} or more modern texts like \cite{press2007mcmc} for a detailed description. In short, random samples of the model's parameters are drawn around previous states, and acceptance follows a straightforward rule: if the proposed state improves the global likelihood of the data given the model, it is always accepted. If the new state is worse, a statistical acceptance criterion is used, based on the likelihood ratio between the new and the previous states. Over a long chain of iterations, this process yields reliable estimates of the parameters distribution. These distributions can be visualised as joint probability distributions (corner plots), as shown in Figure\,\ref{fig:mcmc}. The derived values for the parameters and the corresponding priors are listed in Table\,\ref{tab:03}. 

Given that uncertainties in the data points are an important aspect of parameter optimisation, we also account for additional factors beyond those quoted in Table\,\ref{tab:02}, which lists the inherent (instrumental/photometric) uncertainties associated with the measurements. In total, we consider three sources of error: (1) inherent uncertainty (Table\,\ref{tab:02}), (2) a systematic effect due to orbital modulation (5\%), and (3) uncertainty in the reddening to the source (Section\,\ref{av}). The total error is calculated as the quadrature sum of these contributions. For the missing values of Table\,\ref{tab:02}, we considered inherent uncertainties based on instrument information and comparable studies in the literature, assigning 2\% for HST/UV and 1\% for HST/\textit{U},\textit{B},\textit{V},\textit{R} and CTIO/\textit{B},\textit{V},\textit{R},\textit{I}.

\begin{table}
\caption{Priors and posteriors for the disk$+$secondary fitting.\label{tab:03}}
\centering
\begin{threeparttable}
{\renewcommand{\arraystretch}{1.25}
\begin{tabular}{c c c}
\hline\hline
Parameter & Prior & Posterior\\
\hline
$M_{\text{irr}}$ [M$_{\odot}$]             & $\mathcal{N}$(1.05,0.30) & 1.05$^{+0.22}_{-0.21}$ \\
$R_{\text{irr}}$ [6\,$\times$\,10$^8$\,cm] & $\mathcal{N}$(1.20,0.85) & 1.12$^{+0.56}_{-0.37}$ \\
$L_{\text{irr}}$ [log erg\,s$^{-1}$]       & $\mathcal{N}$(37.6,0.65) & 37.58$^{+0.31}_{-0.40}$ \\
$A$                                       & $\mathcal{N}$(0.25,0.45) & 0.19$^{+0.18}_{-0.17}$ \\
$i$ [$^{\circ}$]                          & $\mathcal{N}$(70.0,11.0) & 72.14$^{+7.47}_{-7.91}$ \\
$R_{\text{in}}$ [R$_{\text{WD}}$]         & $\mathcal{N}$(1.20,1.05) & 1.10$^{+0.64}_{-0}$ \\
$R_{\text{out}}$ [R$_{\odot}$]            & $\mathcal{N}$(1.60,0.50) & 1.57$^{+0.21}_{-0.25}$ \\
Atmosphere factor                         & $\mathcal{N}$(1.50,0.65) & 1.53$^{+0.37}_{-0.45}$ \\
\hline
\end{tabular}
}
\begin{tablenotes}
\item \small \textbf{Notes.} Priors are expressed as mean value and standard deviation. The disk$+$secondary SED model presented in Figure\,\ref{fig:sed} is built from the posterior values. Errors of $+0$ or $-0$ indicate that the parameter reached the allowed hard limit. Subscripts ``irr'' denote that the parameters correspond to the central irradiating source in the SED fit, which is distinct from the component fitted to the observed X-ray spectrum. 
\end{tablenotes}
\end{threeparttable}
\end{table}

\begin{figure*}
% \centering
	\includegraphics[width=\textwidth]{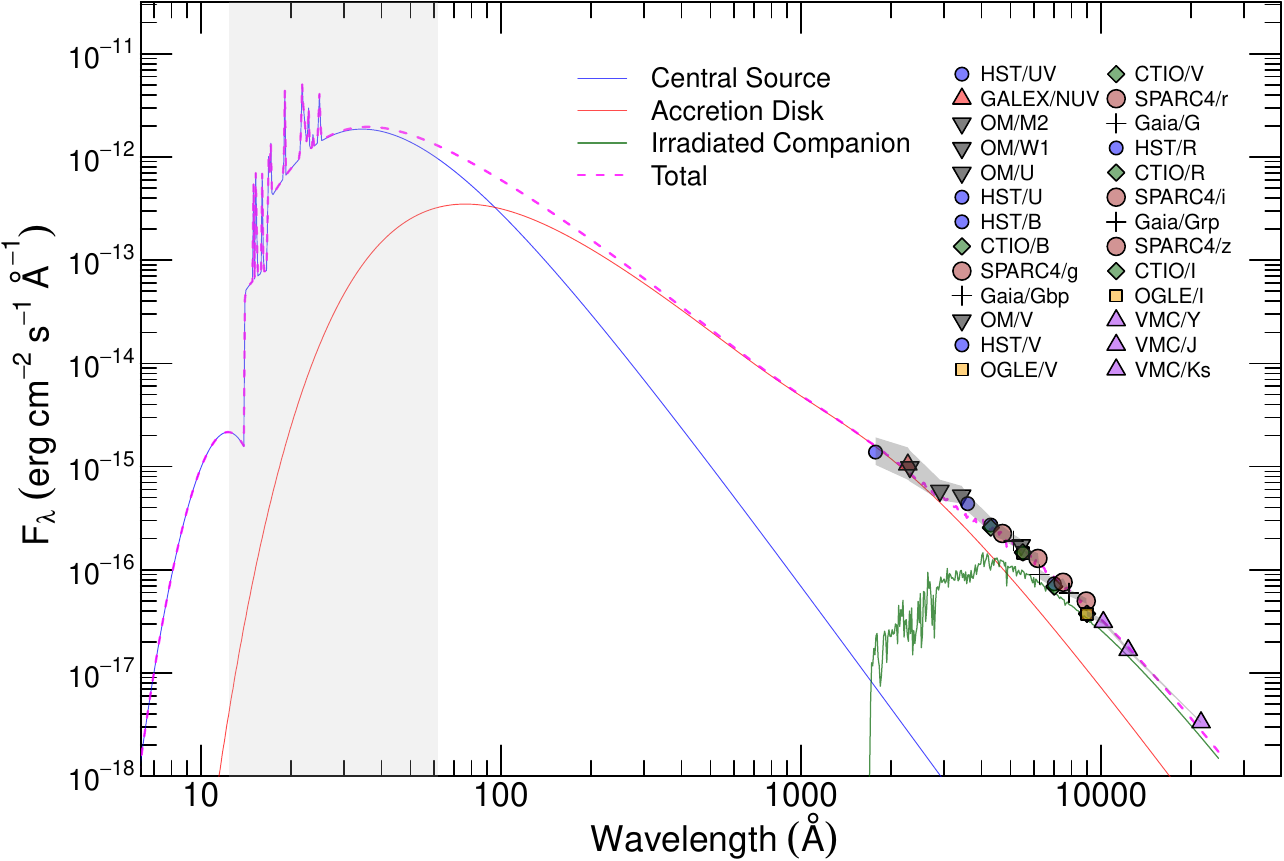}
    \caption{Spectral energy distribution of CAL\,87. The model spectrum representing the central source is derived from the fit to RGS+pn data, corrected for the best-fitting interstellar absorption (i.e. de-absorbed; $N_{\text{H}}$\,=\,0.37\,$\times$\,10$^{22}$\,cm$^{-2}$), and extrapolated to longer wavelengths. The X-ray analysis range (0.2--1.0\,keV) is highlighted in grey. The observed X-ray flux is presumed to arise from scattered emission and is used to guide the prior on the luminosity of the central source, which is varied independently in the SED fitting. The accretion disk was computed using parameters listed in Table\,3. For the irradiated companion, a 7000\,K main-sequence stellar atmosphere model is employed. Measurements at longer wavelengths are dereddened with $A_{\text{V}}$\,=\,0.338$^{+0.144}_{-0.123}$, with shading around the symbols indicating lower and upper limits.}
    \label{fig:sed}
\end{figure*}

\begin{figure*}
% \centering
	\includegraphics[width=\textwidth]{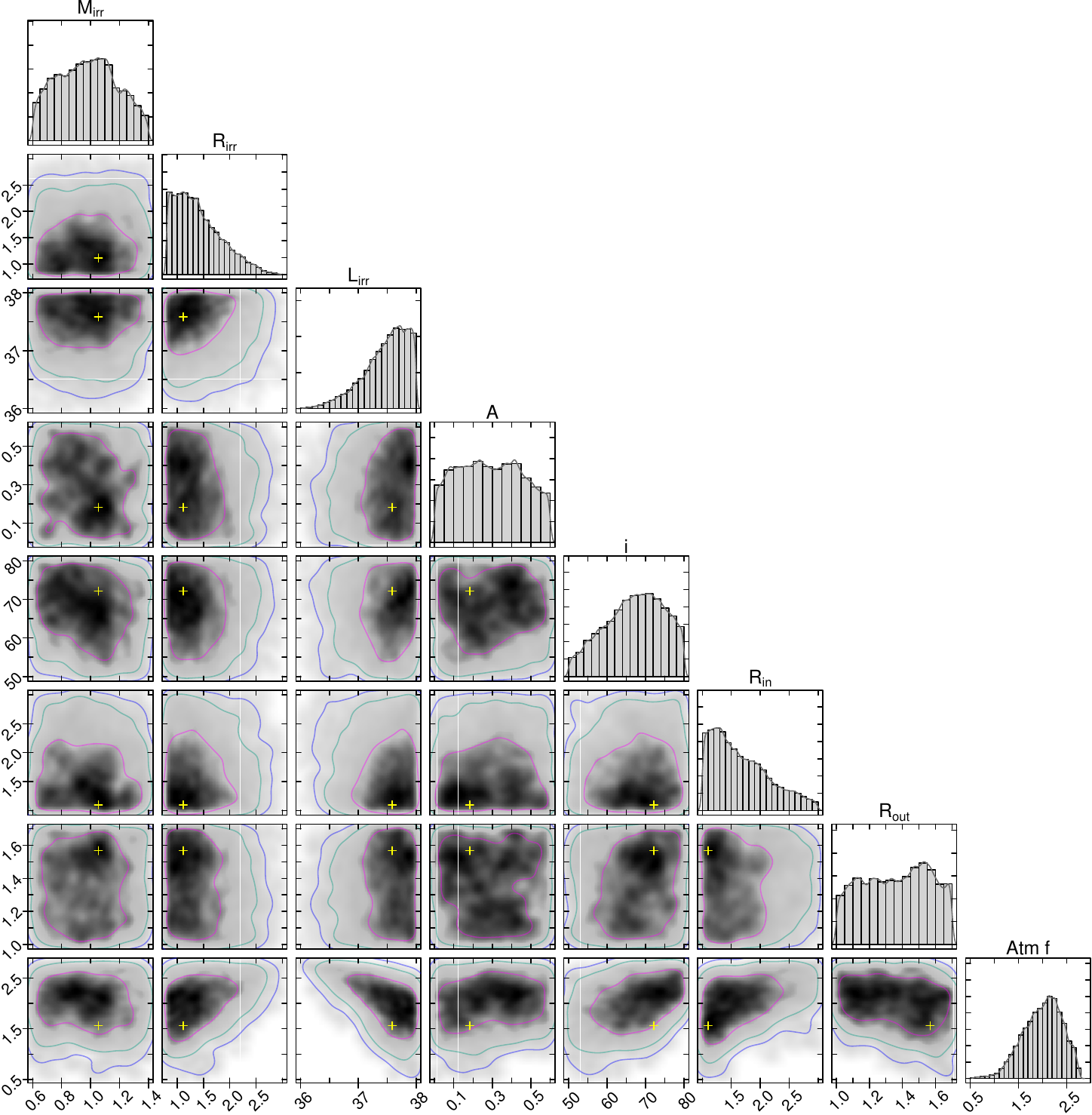}
    \caption{Joint distribution of selected parameters from the accretion disk $+$ irradiated companion model, calculated from the MCMC output. The yellow plus symbols indicate the global best-fitting parameter set, which was used to build the SED model displayed in Figure\,\ref{fig:sed}. Confidence contours are shown for 1\,$\sigma$ (magenta), 2\,$\sigma$ (bluish green) and 3\,$\sigma$ (blue). The parameters and their units are: $M_{\text{irr}}$ (M$_{\odot}$), $R_{\text{irr}}$ (6\,$\times$\,10$^8$\,cm), $L_{\text{irr}}$ (log erg\,s$^{-1}$), $A$ (dimensionless), $i$ ($^{\circ}$), $R_{\text{in}}$ (R$_{\text{WD}}$), $R_{\text{out}}$ (R$_{\odot}$), Atm\,f (dimensionless).       }
    \label{fig:mcmc}
\end{figure*}

\section{Discussion} \label{sec:discussion}

To provide the soft X-ray spectral shape for the SED, we modelled the X-ray spectrum of CAL 87 with a phenomenological model fit simultaneously to the RGS and pn spectra (Section~\ref{sec:dataxray}; Fig.~\ref{fig:xray}). The fit also sets an observed flux scale that served as a broad prior for the central-source luminosity.

Earlier X-ray studies reported significantly lower blackbody temperatures CAL\,87, ranging from 30 to 50 eV (e.g. \citealp{1993PASP..105..863S}; \citealp{1994A&A...288..538K}; \citealp{1998ApJ...503L.143A}). This discrepancy was likely due to the low spectral resolution at the time, which hid the true underlying spectrum by failing to resolve features such as absorption edges and emission lines. As noted by \cite{2001ApJ...550.1007E}, the presence of O\,\textsc{vii} and O\,\textsc{viii} edges requires a temperature between 60 and 80 eV. Our best-fitting temperature, $kT$\,$\sim$\,70--75\,eV, obtained using a model that includes an absorbed blackbody, two edges, and 12 narrow Gaussians at the measured RGS line centroids \citep{2024A&A...690A...9P}, aligns well with this expected range. 

Different values for the X-ray luminosity of CAL\,87 have been reported over the years, reflecting differences in instrumental resolution and spectral modelling, which can affect the inferred absorption column. Reported values typically span from $\sim$\,5\,$\times$\,10$^{36}$ to 10$^{38}$\,erg\,s$^{-1}$ (e.g. \citealp{1997A&A...322..591H}; \citealp{2001ApJ...550.1007E}; \citealp{2024ApJ...960...46T}; \citealp{2024A&A...690A...9P}). Although many of these analyses yield relatively modest values for the unabsorbed luminosity, they often acknowledge that the absolute intrinsic luminosity of the white dwarf may be substantially higher, as the central source is likely obscured by scattering material (e.g. \citealp{2001ApJ...550.1007E}; \citealp{2024A&A...690A...9P}). Our results support an intrinsic luminosity above 10$^{37}$\,erg\,s$^{-1}$. The bolometric luminosity inferred from our SED model is approximately 3.8\,$\times$\,10$^{37}$\,erg\,s$^{-1}$ (Table\,\ref{tab:03}), consistent with expectations for steady nuclear burning in a supersoft source. 

In the spectral energy distribution analysis, we began with the expected---or at least the suspected---value ranges for CAL\,87 as priors. The resulting posteriors allowed us to build a SED model that closely reflects the observed measurements from the UV to the infrared. The WD mass range, 0.84\,$\lesssim$\,$M$(M$_{\odot}$)\,$\lesssim$\,1.27, is consistent with the value of 1.2\,M$_{\odot}$ recently reported by \cite{2024A&A...690A...9P}, which was estimated based on the temperature of approximately 800\,kK derived from four different combinations of atmospheric and plasma models applied to \textit{Chandra} and \textit{XMM$-$Newton} grating spectra. The inclination of the system we find, approximately 72 degrees, falls within previously inferred or considered values (e.g. \citealp{2001ApJ...550.1007E}; \citealp{2007A&A...472L..21O}; \citealp{2014ApJ...792...20R}). 

The literature mentions of the WD radius of CAL\,87 are indirectly based on its probable mass and the mass--radius relationship of \cite{1972ApJ...175..417N}, which we chose not to strictly adhere to, as the WD photosphere should be expanded. Indeed, our SED modelling suggests an effective radius 15--20\% larger than that of a theoretical 0\,K 1\,M$_{\odot}$ WD, consistent with a  photospheric expansion. For the disk radii, the prior for the inner radius was set close to the WD radius, and the posterior converged to a value only slightly larger (about 10\%). The outer radius, meanwhile, was approximately constrained by the primary Roche lobe, assuming a mass ratio $q$\,$<$\,1 (e.g. \citealp{1998ApJ...502..408H}, \citealp{2007A&A...472L..21O}, \citealp{2024MNRAS.527.8991S}). The temperature estimation for the irradiated side of the companion, approximately 7000\,K, derived from $B-V$ colour analysis during the secondary eclipse, complements the overall multi-component SED model. While such equivalent temperature values should be treated with caution, it is also worth noting that our SED modelling clearly indicates the need of a cool source that surpass the irradiated disk model emission in the IR. An extended electron-scattering region (e.g. disk wind/corona or Thomson-scattering halo) around the WD can scatter the emission and reduce shielding by the outer disk, allowing irradiation of the companion. There may be unaccounted contributions to the flux in the IR, which possibly
degenerate with the scaled luminosity of the irradiated secondary found in the multi-parametric fitting. Using an even temperature of 7000\,K over the illuminated hemisphere and the scaled luminosity, one finds a radius that is much larger than the secondary Roche lobe. On the other hand, high temperature gradients are expected between the vicinity of L1 and the terminator (e.g. the illumination modelling of the \mbox{Sco X$-$1} optical light curve by \citealp{2021MNRAS.508.1389C}), which would lead to a smaller radius. 

A scenario with mass loss and an extended electron-scattering region associated with a disk wind/corona has been proposed for CAL\,87. High-resolution spectroscopy (\citealp{2004RMxAC..20..210O}; \citealp{2010AN....331..152E}) and eclipse-reconstructed images in X-rays \citep{2014ApJ...792...20R} provide consistent evidence for an ionised outflow from the disk and/or the central source. Due to the distance to CAL\,87, a circumbinary ejecta would be difficult to resolve. However, \citet{2022AJ....164..145S} claims the presence of nebular continuum emission from a hot wind in 3 LMC/SMC SSS, indicating significant values of mass loss rate. In a constant velocity symmetric wind, the scattered flux from the WD is expected to decay as $r^{-2}$, yielding an extended yet centrally concentrated source. The ratio between the scattered component and the expected steady burning luminosities may be achieved by supposing plausible wind velocities and mass loss rates. However, a detailed wind modelling is required to constrain those quantities. 

We do not assume a unique geometry for the scattering medium. A Thomson-scattering halo at larger radii---invoked for U\,Sco (e.g. \citealp{2012ApJ...745...43N}; \citealp{2013MNRAS.429.1342O})---is an equally plausible alternative to a disk wind/corona for CAL\,87. The modest hardness/continuum changes across eclipse reported by \citet{2014ApJ...792...20R} are consistent with an extended scattered component plus a more compact, partially occulted one. Our RGS+pn spectra do not discriminate between these cases, and the SED is insensitive to this choice because the irradiating luminosity is fitted independently of the X-ray flux.

In this study, a multiwavelength modelling of the basic system components is made. The amount of data currently available allow the assembling of an averaged and wavelength-extended SED, which provides a robust estimate of the main source luminosity and other system parameters. Precision is compromised in the process of combining data from different epochs, although only out-of-eclipse points were selected at the same photometric high-state. Extended low-states have not been observed in this source so far. With relatively low distance and $N_{\text{H}}$ uncertainties, the luminosity of the observed scattering source is mainly defined by the X-ray spectrum fit, while the UV, optical and NIR data mostly concern the accretion disk and companion star emission properties. Unfortunately, the usual data gap between soft X-rays and UV hides the most discriminant information regarding the inner accretion disk. The multi-parametric SED fitting suggests a range of orbital inclinations that are consistent with non-total primary eclipses of an extended accretion disk. Given the estimated WD mass, CAL\,87 would be in the steady-burning regime for an accretion rate of $\dot{M}\sim\text{1--3}\times10^{-7}$\,M$_\odot$\,yr$^{-1}$ (see \citealp{2013ApJ...777..136W}). These values are consistent with the derived X-ray luminosity under the assumption of long-term equilibrium burning of H-rich accreted material. 

The irradiated disk luminosity is constrained to 10$^{37}$\,erg\,s$^{-1}$. Our model excludes any contribution from viscous dissipation in the disk. Such a contribution was found to be smaller than 7\% of the disk total luminosity for $M_{\text{WD}}$\,$\times$\,$\dot{M}$\,$\lesssim$\,$10^{-7}$\,M$_{\odot}^2$\,yr$^{-1}$, considering the WD luminosity estimated from the SED modelling. The emission from a standard viscous disk with a multi-temperature blackbody SED, across a range of plausible mass accretion rates for CAL\,87, would be significantly weaker than that of the computed irradiated disk in the EUV region. At other wavelengths shown in Figure\,\ref{fig:sed},  the composite SED modelled here would far outshine a non-irradiated viscous disk. In a combined viscous plus irradiated disk emission model, the relative contribution of the inner, high $\dot{M}$ disk dissipation in the EUV may eventually disentangle the degeneracy between $M_{\text{WD}}$\,$\times$\,$\dot{M}$ and the WD photosphere irradiating luminosity. Therefore, further measuring of the SED in the EUV seems important to probe the viscous disk structure and refine constraints on the mass transfer rate.

\section{Conclusion}

A comprehensive analysis of the spectral energy distribution of the supersoft X-ray source CAL\,87 is presented. Careful attention was given to retrieve robust data from out-of-eclipse phases and combine them with proper calibration. Spanning wavelengths from X-rays to the near-infrared, the revised SED provides constraints for the multiple radiation sources in the system. 

By employing a multi-component model, including a reprocessed accretion disk and an irradiated companion atmosphere, we have achieved a close match to the observed fluxes. The revised parameters, including the central source total luminosity, are well constrained by the improved sampling and coverage of the SED. They proved to be highly plausible for this system while many of them are supported by previous analyses. The equivalent temperature for the eclipsed companion photosphere based on its $B-V$ colour during the secondary eclipse indicates that it is the main IR source in the system.

Future work could involve testing improved models for the central source, accretion disk and companion emission using additional observations at different wavelengths, especially in the EUV. The SED analysis suggests that a narrower EUV gap is essential for probing the disk viscous dissipation and accretion rate. On the other hand, phase-resolved optical/infrared high-resolution spectra may reveal emission line components from the companion irradiated atmosphere. Furthermore, extending the up-to-date SED analysis to other supersoft X-ray sources may define a broader context for comparing systems like CAL\,87.






% \bibitem[Lamport(1994)]{lamport94}
%   \textit{\LaTeX: a document preparation system},
%   Addison Wesley, Massachusetts,

% \end{thebibliography}

\section*{Data availability}
Data will be made available on request.

\section*{Acknowledgments}
MPD thanks Conselho Nacional de Desenvolvimento Científico e Tecnológico (CNPq) under grant 310309. ASO acknowledges São Paulo Research Foundation (FAPESP) for financial support under grant no. 2017/20309-7. NP acknowledges that this study was financed in part by the Coordenação de Aperfeiçoamento de Pessoal de Nível Superior - Brasil (CAPES) - Finance Code 001. FJ and FD thank the Brazilian Ministry of Science, Technology and Innovation (MCTI) and the Brazilian Space Agency (AEB), who supported the present work under the \mbox{PO\,20VB.0009}

\bibliographystyle{elsarticle-harv} 
\bibliography{cal87_sed}

\end{document}